# Versatile Hall magnetometer with variable sensitivity assembly for characterization of the magnetic properties of nanoparticles

Jefferson F. D. F. Araujo[1]*, Daniel R. P. Vieira[1], Fredy Osorio[1], Walmir E. Pöttker [2], Felipe A. La Porta[2], Patricia de la Presa [3,4], Geronimo Perez [5] and Antonio C. Bruno[1]

[1]*Physics Department, Pontifical Catholic University of Rio de Janeiro, Rio de Janeiro 22451-900, Brazil*

[2]*Federal Technological University of Paraná, Laboratory of Nanotechnology and Computational Chemistry, Avenida dos Pioneiros 3131, 86036-370, Londrina, PR, Brazil*

[3]*Institute of Applied Magnetism, UCM-ADIF-CSIC, A6 22,500km, 28230 Las Rozas, Spain*

[4]*Material Physics Department, UCM, Ciudad Universitaria, 28040 Madrid, Spain*

[5]*Materials Metrology Division, National Institute of Metrology, Standardization and IndustrialQuality (Inmetro), Av. Nossa Senhora das Graças, 50, Xerém, CEP: 25250-020, Duque de Caxias, RJ, Brazil*

*Corresponding author: jferraz@fis.puc-rio.br*

**Abstract** − A Hall magnetometer with variable sensitivity is constructed to measure the magnetic properties of magnetic nanoparticles manufactured by different methods. This novel magnetometer can also be used to measure bulk materials and samples in liquids. The magnetometer is constructed with two commercial Hall-effect sensors in an acrylic structure, which serves as the support for a micrometer and the circuit board with the sensors. For operation, the magnetometer it acquires a complete magnetization curve in a few minutes. If has a magnetic moment sensitivity of $1.3 \times 10^{-9}$ Am$^2$ to sensitivity of 493 mV/mT, the sensitivity can be adjustable in the range of 10 to 493 mV/mT. Its performance is tested with magnetic nanoparticles. As an application example, we estimate the mean diameter of the nanoparticles using the magnetic curves. The results are compared with those obtained by other techniques, such as transmission electron microscopy (TEM), X-ray diffraction (XRD) and dynamic light scattering (DLS). The magnetization results are also compared with those obtained by independent commercial magnetometers, which reveals errors of approximately ±0.31 Am$^2$/kg (i.e., 0.6%) in the saturation region.

# 1. Introduction

Extensive studies on magnetic properties of diverse advanced materials have been performed to analyse physical phenomena and evaluate their potentials for engineering applications [1-5]. The Curie temperature, saturation magnetization, remanent magnetization, coercive field, and magnetic anisotropy are among the most important parameters of magnetic materials, determining their applicability. Several magnetometers, including those based on a superconducting quantum interference device (SQUID) [6], magneto-optical Kerr effect [7-8], alternating gradient [9], vibrating samples (VSM) [10], and the Hall effect [11,12], with different sensitivities and costs, have been designed and successfully used. Also, magnetic nanoparticles have a large potential for biomedical applications [13-15]. With the progress in the applications of magnetic materials in several fields of nanotechnology, particularly in the last decades, the interest in magnetometers increases. The extensive studies have led to the development of high-quality devices with a wide range of sensitivities.

However, most of these magnetometers are expensive and thus unaffordable for use in low-budget laboratories. In this study, we present a different measurement approach for the magnetization curves using a magnetometer based on Hall-effect sensors with variable sensitivities. It enables to measure in a cylindrical format (sample holder in cylindrical format), samples bulk, liquids, microparticles and nanoparticles with mass in the range approximately 900 µg and 40 mg without the need to change sensors or measurement systems. In this system, we use equipment normally accessible to most laboratories, including an electromagnet, current source, and data acquisition module. A particular advantage of this approach is that the magnetometer can be rapidly assembled and disassembled. Therefore, the equipment can also be used in other experimental configurations. The core of the magnetometer consists of an acrylic plate, which acts as a support for the sensors, micrometer, and the sample holder.

In most magnetometers, a magnetic dipole model is usually used to obtain the moment of the sample. However, when the sample is near the sensor, as in the assembled magnetometer, the sample geometry cannot be neglected. Therefore, the dipole model commonly used to obtain the magnetic moment of the sample introduces significant errors [16]. The model is improved by considering the geometry of the sample holder to more accurately estimate the magnetic moment of the sample. We test the magnetometer with magnetic nanoparticles fabricated using the co-precipitation method. This method is fast and versatile, and allows the obtaining of a great variety of magnetic nanoparticles presenting a great number of advantages, such as small reaction time, obtaining particles with few agglomerations, low cost, large quantities and still allows us to control size and distribution of the obtained nanoparticles, through the variation of parameters such as pH [1,11-12]. The mean diameter of the magnetic cores of the nanoparticles is estimated through the magnetic curves, and compared with those obtained using other techniques, such as transmission electron microscopy (TEM), X-ray diffraction (XRD) and dynamic light scattering (DLS). The prepared samples weigh a few tens of milligrams.

## 2. Magnetometer probe design

There are only a few basic components used in the structure of the magnetometer the electromagnet, direct current (DC) source, and Hall-effect sensor. All of the parts necessary to assemble the magnetometer are shown in Fig. 1, which include the following components: (a) acrylic structure (200 mm (length) x 72 mm (width) with slot dimensions of 86 x 45 mm$^2$), which serves as a support for the micrometer and as the base where the sensors are mounted; (b) a micrometer, which aligns sensor 1 with respect to the magnetic field applied by the electromagnet; (c) a cylindrical arm (14 mm (diameter) x 75 mm (length)), connected to the micrometer; (d) acrylic piece (45 x 45 x 22 mm$^3$), which is connected to the arm and has two perpendicular grooves, one of them is used to fasten the Hall devices (45 x 8 mm$^2$) connected to the cylindrical arm, while the other is used for the sample holder; (e) circuit board (8 x 32 x 2 mm$^3$), where two Hall-effect sensors (AD-AD22151, Analog Devices, Inc.) are mounted; (f) acrylic piece (5.0 x 3.0 x 30 mm$^3$), which is the sample holder used to accommodate samples in a cylindrical cavity with a diameter of 3.0 mm and length of 3.0 mm; and (g) the poles of the electromagnet (diameter: 40 mm).

The sample sensor is connected to the micrometer at the cylindrical arm (see Fig. 1.(c)), so that it can be rotated and placed as perpendicular as possible to the electromagnet before each measurement. In this setting, the sensor senses as low as possible magnitude of the electromagnet's applied field. This system is placed between the poles of the electromagnet and connected to a data acquisition board, through a filter (SR560 low-noise preamplifier, Stanford Research Systems), and to a computer. A program was written in LabView® to control the current source and receive data from the two sensors (in volts) during the measurement procedure.

A voltage of 5.0 V DC was applied to circuit; under these conditions, the sensors have adjustable sensitivities of 10 to 493 mV/mT. The sensor has an active area of 0.20 x 0.20 mm$^2$. As shown in Fig. 1, the magnetometer was tested by attaching it to an acrylic plate between the poles of the electromagnet (3470 GMW, Inc.). This plate contains two AD22151 Hall-effect sensors (sensor 1 measures the induced field of the sample, while sensor 2 measures the field applied by the electromagnet). With a pole diameter of 40-mm, this electromagnet can generate a field of 1.0 T at a pole gap of 10 mm and current of 5.0 A. Using our magnetometer probe, an entire magnetization curve, with approximately 60 points, is generated in less than 15 min. A thermocouple connected to the pole at the sensor position detected no change in temperature during the measurement period, even at 1.0 T. However, the AD22151 Hall sensor has a sensitivity drift error smaller than +1% over a wide temperature range. We retained the sensitivity temperature compensation setting at its factory default, which is typically -500 ppm/°C; no water cooling is required to operate the electromagnet at this level. In addition, at this pole gap (width of the acrylic plate), the generated magnetic field is homogeneous, which the homogeneity is suitable for the as-prepared sample sizes.

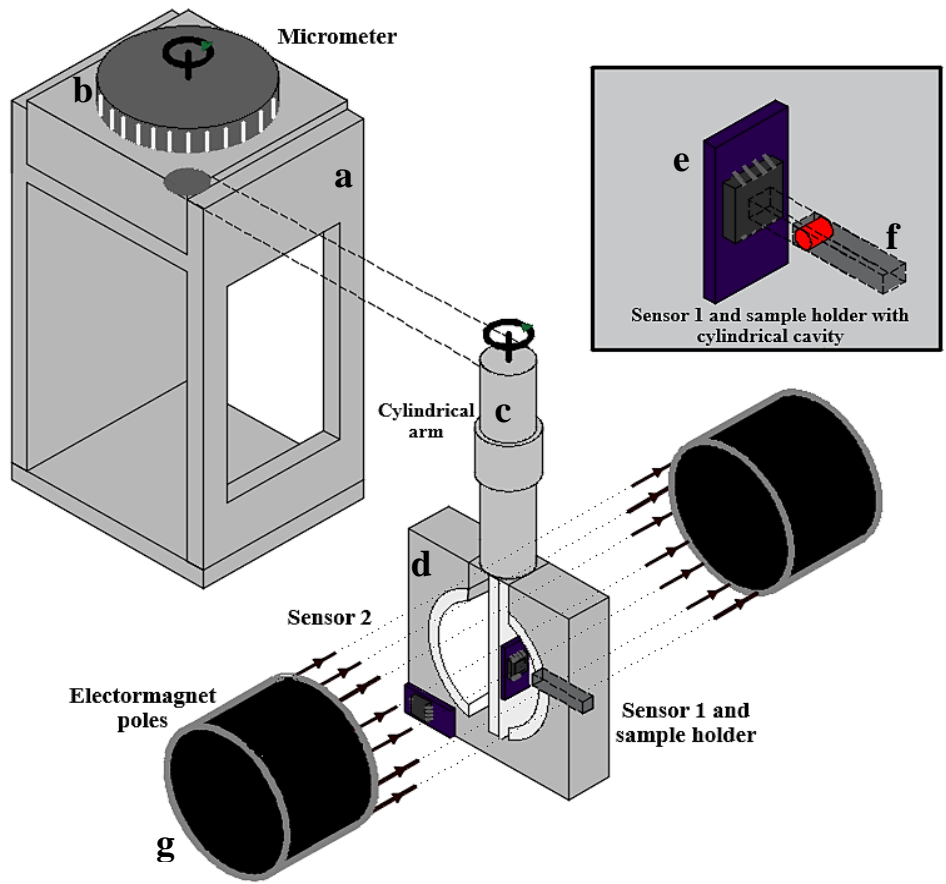

**Fig. 1.** Components of the constructed magnetometer: (a) Acrylic structure [200 mm (length) x 72 mm (width) with slot dimensions of 86 x 45 mm$^2$]; (b) Micrometer; (c) Cylindrical arm [14 mm (diameter) x 75 mm (length)]; (d) Acrylic piece (45 x 45 x 25 mm$^3$) connected to the arm; (e) Circuit board (8 x 32 x 2 mm$^3$), where two AD22151 Hall-effect sensors are mounted; (f) Acrylic sample holder (5.0 x 3.0 x 30 mm$^3$); (g) electromagnet pole (diameter: 40 mm).

The internal structure of the AD22151 Hall sensor a contains Wheatstone bridge, where the sensitivity values are related to the resistances used to close the operating circuit, as indicated in Table 1. The resistors are placed on a board connected to the acquisition unit (model 34970A, Agilent), where the user can select the resistors used to close the circuit. This resistance change procedure using the acquisition unit was automated and coupled in a program written in LabView®. The values shown in the second column of Table 1 are the results of the calibration of the Hall sensors, i.e., the results of the analysis used to determine the voltage produced when the sensor responds to a field of a known magnitude. Both sensors are calibrated using a pair of Helmholtz coils. In order to measure the magnetic field of the pair of coils, we use a Gaussmeter model 5080, produced by F.W. Bell.

Table 1. Available resistances, corresponding sensitivities, and signal-to-noise ratios.

| Resistance (kΩ) | Sensitivity (mV/mT) | Signal - To – Noise Ratio |
|---|---|---|
| 1 | 13.8 | 148 |
| 10 | 55.2 | 135 |
| 22 | 119 | 131 |
| 47 | 224 | 142 |
| 100 | 493 | 138 |

Table 1 shows the measured signal-to-noise ratios, which are approximately 140 for all of the available sensitivities. Therefore, if necessary, the sensitivity can be increased by keeping the signal-to-noise ratio approximately constant. These values are calculated using a Stanford Research Systems spectrum analyzer model SR760. Fig. 2.(a) shows results of a measurement performed using a spectrum analyzer, which can be used to compare the sensitivities. The data show that the signal-to-noise ratio for the lowest resistance is slightly lower than the others. In Fig. 2.(b), the signal-to-noise ratio of the sensor used in our self-built magnetometer, AD22151, is compared to that of a commercial sensor, MLX90215 [11]. Although the Hall sensor model MLX90215 has programmable sensitivities between 3.1 and 140 mV/mT, it is difficult to change the sensitivity as it is necessary to separate the sensor from the piece and program it on other equipment. This implies that the entire system must be disassembled, including the sensor from the acrylic plate, which may damage the sensor. In the case of our magnetometer, it is not necessary to disassemble the configuration to vary the sensitivity. Therefore, the new system is advantageous with respect to sensitivity changes and disassembly method.

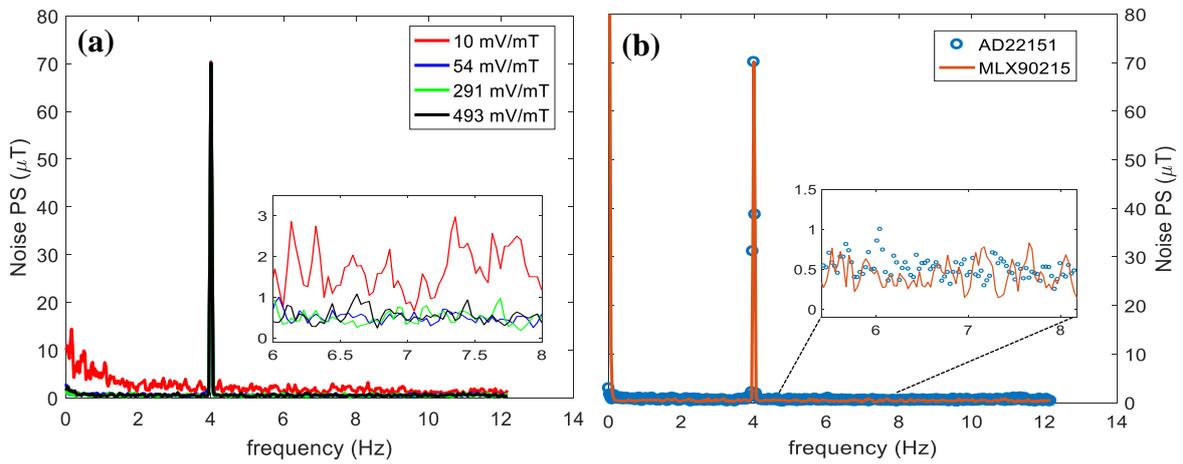

**Fig. 2.** (a) Signal-to-noise ratio graph for a 4 Hz signal at sensitivities of 10 to 493 mV/mT; the magnified view shows a comparison of noises in the range of 6 to 8 Hz. (b) Comparative signal-to-noise ratio graph of the two sensors; the magnified view shows a comparison of noises in the range of 5 to 6.5 Hz.

The final assembly is shown in Fig. 3.(a) and Fig. 3.(c), whereas the sample holder with the nickel sphere used to calibrate the Hall magnetometer is shown in Fig. 3.(b). The figure also shows the circuit assembly and Hall sensor just below the sample holder. The Hall sensor is an AD22151 linear magnetic field transducer manufactured by Analog Devices. This device reads a voltage proportional to the magnetic field applied perpendicularly. The bipolar configuration with a low compensation was used (< -500 ppm, AD-AD22151, Analog Devices, Inc.).

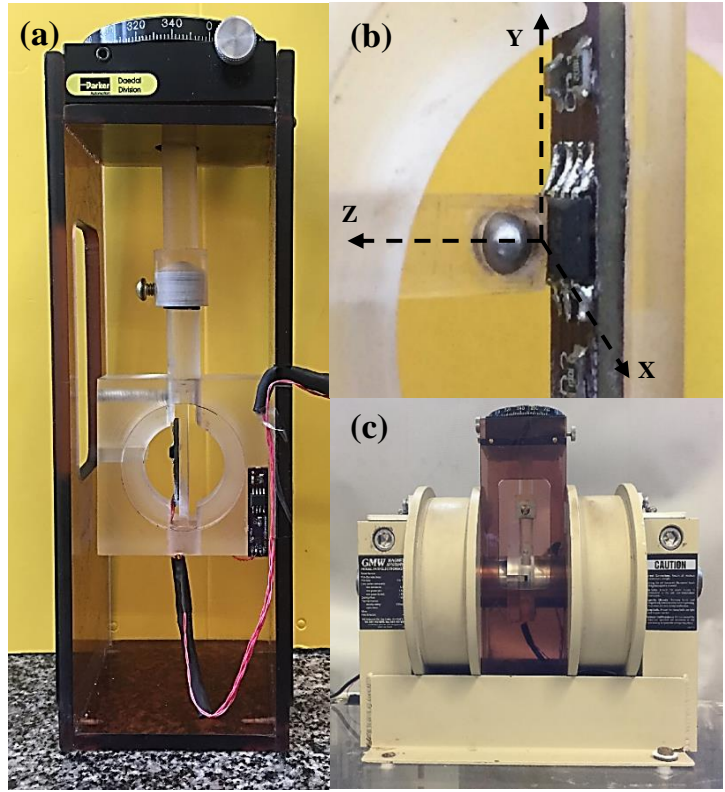

**Fig. 3.** (a) Final assembly of the acrylic structure, including the part that serves as the support for the micrometer and base where the sensors are mounted on the acrylic piece. (b) Acrylic piece connected to the cylindrical arm and sample holder with a nickel sphere positioned exactly above the AD22151 Hall sensor, which was installed on the circuit board, the axes indicate the x-, y- and z-directions. (c) Assembled magnetometer.

The AD22151 consists of n-EPI Hall plate structures located in the center of the sensor. Therefore, the signals from the transducer pass through commutation switches in an orthogonal sampling process through a differential amplifier (Fig. 4.(a)). From this step, the two generated signals are synchronously demodulated to cancel the resultant offset. Finally, the signal is adjusted in a noninverting amplifier and the gain is provided through an external circuit setup. A printed circuit board (PCB) is introduced to physically support the sensor (Fig. 4.(b)), supply power, and modify the gain of the output to calibrate the measurement. The gain can be expressed as:

$$Gain = 1 + \left(\frac{R_2}{R_3}\right) 4 \left[\frac{mV}{mT}\right] \qquad (1)$$

where R$_2$ is 1 kΩ and R$_3$ is a trimmer, located outside the PCB to enable a gain modification.

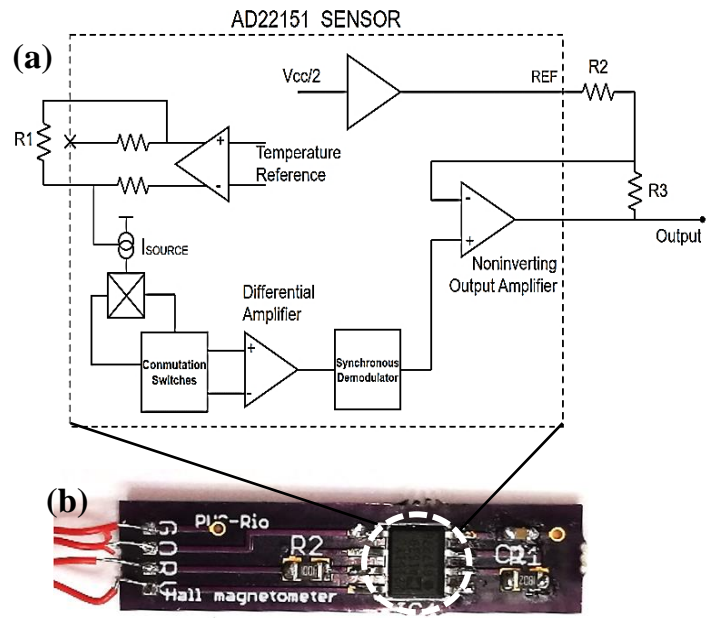

**Fig. 4.** (a) AD22151 internal schematic; (b) Picture of the circuit board.

## 3. Model

### 3.1. Dipole model

The Hall sensors measure the magnetic fields; however, we must obtain the magnetic field normalized by sample volume to characterize the material. The theoretical models used to relate the magnetic field to the magnetic moment require the distance between the sample and the sensor, which is fixed in our equipment design. Using an optical microscope, we evaluated the distances, along both x- and y-axis, from the center of the sample holder to the center of the active area of the sensor. Because the distance along the z-axis could not be directly evaluated, we estimated it by measuring a Ni sphere with known saturation magnetization. Because of the spherical geometry of the sample, the magnetostatic dipole equation can be applied [17]. In Eq. 2, the magnetization in the x-direction and the field measured in the z-direction are considered as follows:

$$\boldsymbol{B}_z(x,y,z) = \frac{\mu_0}{4\pi}\left[\frac{3(m_x x)z}{(x^2+y^2+z^2)^{\frac{5}{2}}}\right] \quad (2)$$

where x, y, and z represent the distances between the sample holder and sensor, $m_x$ is the magnetic moment of the sample and $\mu_0$ is the permeability of free space. The mass of our 99% pure Ni sphere (Goodfellow, Inc.) is 126 mg; we assumed that the mass magnetization for Ni at 1.0 T is 55.14 $Am^2$/kg [18]. Using these values, the resulting z distance is obtained from fitting the measurement to Eq. 2, which is 3,02 mm and Fig. 5 shows this result compared with a measurement in a stand-alone commercial magnetometer (MPMS SQUID, Quantum Design Inc.). They have a relative error of 0.6% in the saturation region and of 1,9 % in the no applied field (remanence) region.

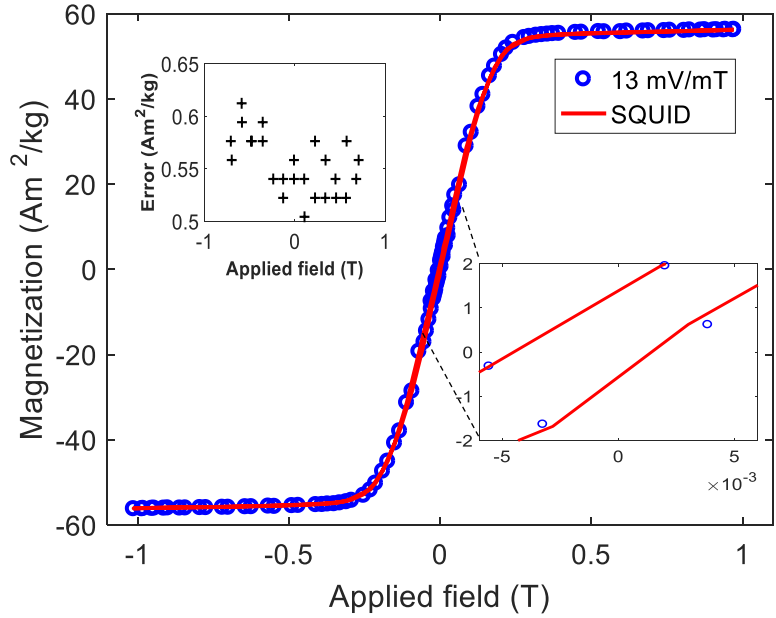

**Fig. 5.** Comparison of the magnetisation curves of the Ni sphere measured with our Hall magnetometer in the 13 mV/mT sensitivity (circles) configuration and with a SQUID magnetometer (solid curve). The lower inset shows that the remanent magnetization obtained by the two magnetometers is less than 0.5 $Am^2$/kg different. The upper inset shows the difference between SQUID and Hall results as a function of the applied field.

### 3.2. Current cylinder model

The dipole model is useful when the sample is far away from the sensor or when the sample is spherical, which is not the case in this study [19]. Considering the cylindrical design of the sample holder, we chose a cylindrical current sheet model to characterize any sample assuming the cavity format. If the cavity has a radius of $a$, length of $L$ and its axis is along the x-direction, the magnetic field in the z-direction can be derived from the Biot-Savart's law. After the calculations, using the definition of the magnetic moment, we obtain [19-20]:

$$B_z(x,y,z) = \frac{\mu_0 m_x}{(4\pi)(\pi a^2)} \int_{-L/2}^{L/2} \int_0^{2\pi} \frac{xa \cos(\phi) d\phi}{r^3} dx_0 \quad (3)$$

where $r = \sqrt{(x)^2 + (y - a\sin(\phi))^2 + (z - a\cos(\phi))^2}$.

In order to test this model (see Eq. 3), we used 19.5 mg of nickel ferrite ($NiFe_2O_4$) nanoparticles that, $NiFe_2O_4$ nanoparticles have been prepared following the reported standard protocol by co-precipitacion of 2.5 mmol of $NiCl_2.6H_2O$ is dissolved in 50 mL of distilled water. On the other hand, 5 mmol of $FeCl_3.6H_2O$ is dissolved in 50 ml of distilled water. Both solutions are heated at 50 ºC and mixed. A solution of 100 ml NaOH (3 M) at 95 ºC was used as precipitating agent. Metal chloride and Sodium hydroxide solutions were added drop wise from two separate burets into a reaction vessel containing 100 mL of distilled water for obtaining uniform particle size distribution [1]. The synthesis temperature is kept constant at 80 ºC for to 120 min under magnetic stirring (600 rpm). After that, the mixture is cooled down to room temperature and magnetically separated, washed several times with distilled water under sonication. In Fig. 6, we can see the measured curve with a sensitivity of 13 mV/mT is compared with that measured using a Hall magnetometer with a fixed sensitivity (58 mV/mT). They well agree, with an error of approximately 1.75%.

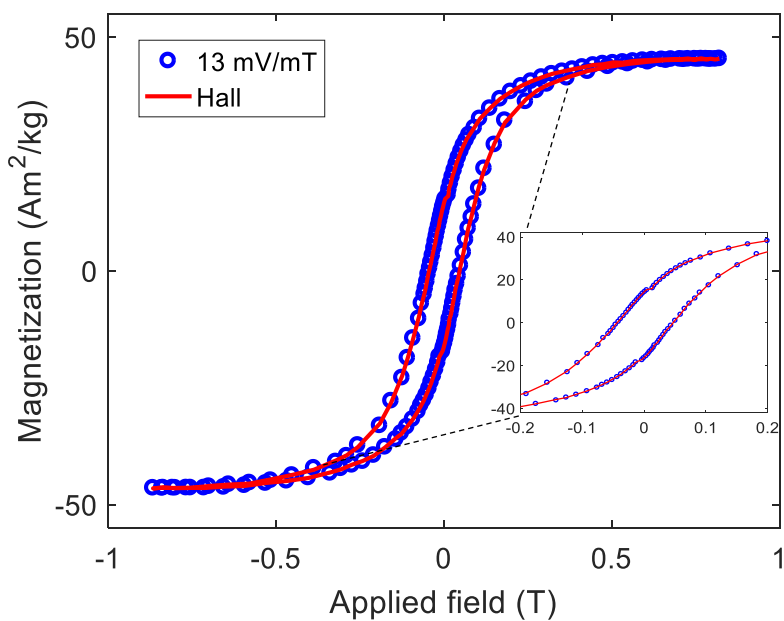

**Fig. 6.** Measurements of NiFe$_2$O$_4$ particles using two different sensitivities. Comparison of the magnetization of Ni fine particles obtained with our magnetometer using the current cylinder model (circles) with the results obtained using the same particles and the Hall magnetometer (solid curve). The inset shows the zero-field region.

**4. Results and discussion**

Samples of cobalt ferrite (CoFe$_2$O$_4$) nanoparticles were synthesized by the co-precipitation method. Nanoparticles of CoFe$_2$O$_4$ were prepared by co-precipitation following the protocol. First, 2.0 mol.L$^{-1}$ solutions of FeCl$_3$.6H$_2$O and 1.0 mol.L$^{-1}$ of CoCl$_2$.6H$_2$O are prepared immediately before the reaction, maintaining the molar ratio 2:1. Both solutions are mixed together. Then NaOH (0.5 M) was added. A volume of 2.5 ml of the salt mixture is injected into 25 ml of an ammonium hydroxide solution (NH$_4$OH). Nanoparticle precipitation occurs at high pH (pH 11-12). This is a material with low magnetic response that was used to test the various sensitivities available in the new magnetometer. This type of sample with low magnetic response cannot be measured in using, for example, a Hall probe magnetometer developed by Araujo and co-workers [11-12]. A first measurement of the 19.5 mg sample was carried out at a sensitivity of 290 mV/mT to ensure reading quality by comparison with results from a VSM model 7404 (Lakeshore Inc.), shown in Fig. 7.(a). Nanoparticles of $CoFe_2O_4$ is a sample with a lower 10x magnetization when compared to the sample of Fig. 6. Unlike this VSM model, our magnetometer has a maximum range of approximately 1.0 T owing to the electromagnet's limitations to withstanding the high current required to generate higher fields. Nevertheless, its accessible range is sufficient to obtain excellent cha racterizations of most materials, such as $CoFe_2O_4$.

A series of magnetic characteristics can be identified from Fig. 7. (a). According to the magnified view of the curve shown in Fig. 7. (b), the remanent magnetization (Mr), i.e., the magnetization of the sample when the field returns to zero after it has reached the cycle's maximum, is approximately 0.4 $Am^2kg^{-1}$, while the coercive field, i.e., the opposite field value necessary to obtain zero magnetization in the sample, is approximately 5.5 mT. These values show that the nanoparticles has a behavior superparamagnetic, i.e., they are characterized by the value of remanence low and therefore have a low coercive field, with no apparent saturation as we can see in Fig. 7.(a). The relative error between the curves is approximately 1.8% (Fig. 7.(b)). Measurements of the same sample at different sensitivities were made to verify the versatility of the device, as shown in Fig. 7.(c). In a low-magnetization material such as cobalt ferrite, the lowest sensitivity (13 mV/mT) produces very noisy readings, whereas an intermediate sensitivity (290 mV/mT) produces a good signal. This finding suggests that even less intense magnetization materials can be measured, or smaller masses of materials with similar or greater magnetic response. This result is important because it indicates that the desired versatility was achieved and that we were able to measure samples with very different magnetization levels. In Fig. 7.(d) we can check the magnetic curve in the region of low applied magnetic field, with two different sensitivities of 290 mV/mT (solid curve in blue color) and 13 mV/mT (circles - graph with repeatability error bars), when the material is demagnetized , this first curve is normally called the first magnetization curve or only the virgin curve.

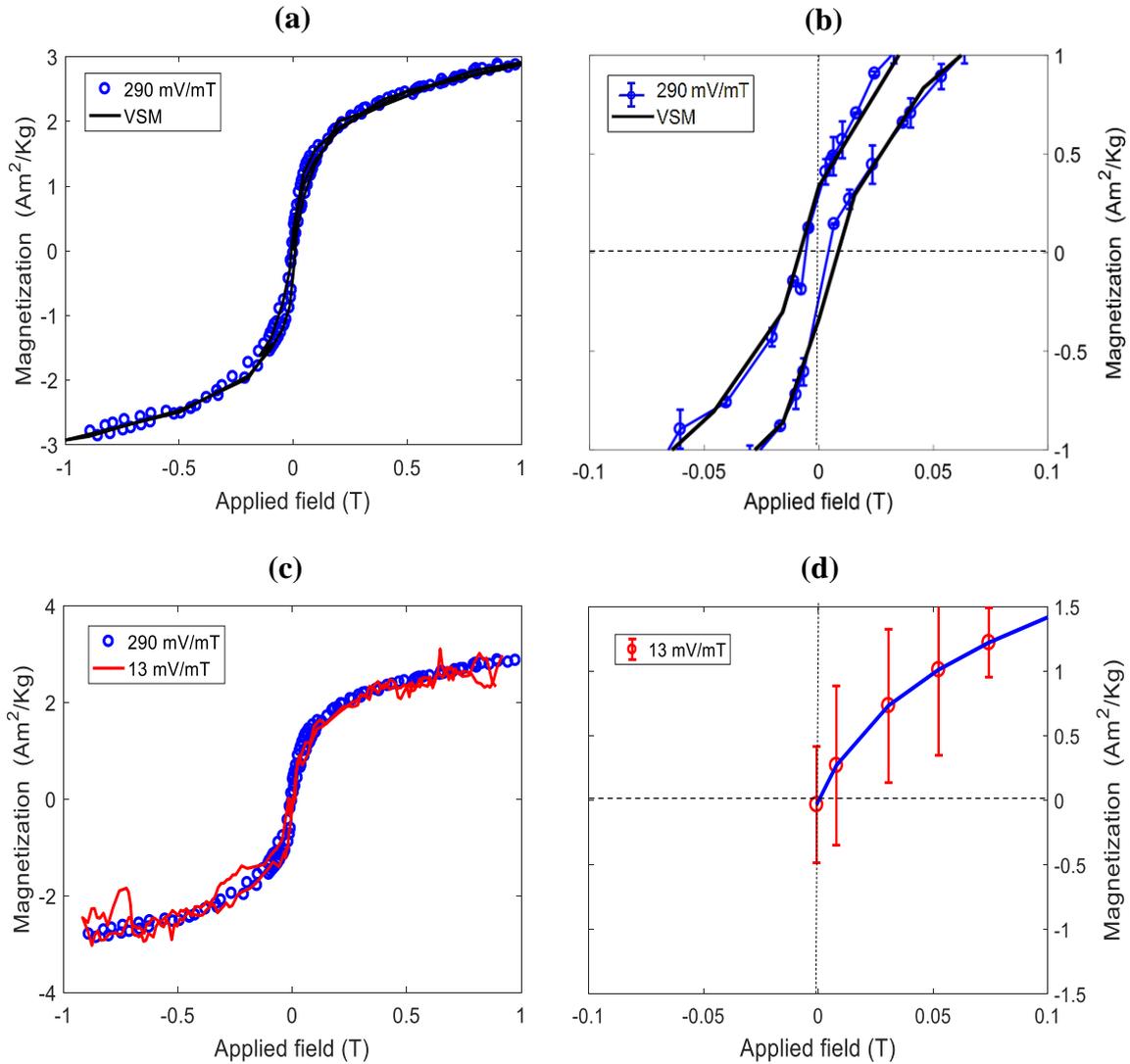

**Fig. 7.** (a) Comparison of magnetization curves of the cobalt ferrite nanoparticles measured with our magnetometer in the 290 mV/mT (circles) configuration and those measured with the VSM magnetometer (solid curve). (b) Zero-field region with the repeatability error bars in the 290 mV/mT (solid curve in the blue color) from the measurements and the measured in the VSM magnetometer (solid curve in the black color). (c) Magnetization curves for cobalt ferrite measured with two different sensitivities of 290 mV/mT (circles) and 13 mV/mT (solid curve). (d) Virgin hysteresis curve in the region of low applied magnetic field, with two different sensitivities of 290 mV/mT (solid curve in blue color) and 13 mV/mT (circles - graph with repeatability error bars).

## 4.1. Characterization of magnetic nanoparticles

Nanoparticles are an important tool in medicine for diagnosis and treatment of various diseases. Their sizes can be controlled, ranging from tens to hundreds of nanometers, which enables them to interact with cells, bacteria, and viruses. Magnetic nanoparticles have a core of magnetic materials. Knowledge of the magnetic properties of nanoparticles is very important not only in the manufacturing process but also in their use. In general, the core of the magnetic nanoparticles are obtaineds using the following techniques, such as TEM, XRD and DLS. We show that it is possible to estimate the thickness of the coating material exclusively from the magnetic measurements.

Nanoparticle diameter can be estimated through evaluation of the magnetization curve close to the origin, which can be deduced from the first approximations of the Langevin function [21]. The derivation of the classical Langevin model can be found in several magnetism books [22], with its result being:

$$\frac{M}{M_S} = L(a) = coth(a) - \frac{1}{a}; a = \frac{\mu H}{k_B T} \qquad (4)$$

In Eq. 4, M is the magnetization, $M_S$ the saturation magnetization, µ the magnetic moment, H the applied field, $k_B$ the Boltzmann constant and T the temperature in Kelvin. This function can be expanded in a series Taylor, in which the first term can be considered a good approximation in low fields or high temperatures as follows:

$$\frac{M}{M_S} \cong \frac{a}{3} = \frac{\mu H}{3k_B T} \qquad (5)$$

This approximation is used in the formula for calculating the diameter. The primary mathematical manipulation is the substitution of magnetization per volume M by magnetization per mass σ using the density of the material ρ:

$$M = \sigma\rho \qquad (6)$$

Substituting Eq. 6 into Eq. 5 will produce σ/H, which can be measured more accurately as the slope of the magnetization curves obtained with the magnetometer ($\chi$ – susceptibility). Since the approximation in Eq. 5 only applies to low fields, we must use the slope near zero, which will be indicated by the value 0 in the slope term. Because the particles are spherical, the magnetization is volumetrically uniform, such that $M_s$ = nµ = µ/$V_{particle}$. Adding the constant µ$_o$ to change the units to the SI system, we estimate the particle diameter as follows [12]:

$$D_{mag} = \frac{(18 k_B T \chi)^{1/3}}{(\rho \pi M_S^2)^{1/3}} \qquad (7)$$

where $D_{mag}$ is the average diameter to be calculated, T is room temperature and $\chi$ is the initial susceptibility.

For the $CoFe_2O_4$ density, we used the value of $\rho = 5.29 \times 10^3$ kg/m$^3$. Using the measurement results (Fig. 7.(d)) for fields around zero, $\chi$ = 211 Am$^2$/kgT. Hence, the $M_s$ is estimated by extrapolating the magnetization curve as a function of the inverse field (1/H); for 1/H = 0, $M_s$ = 13.5 Am$^2$/kg. Using these values, the average diameter of the nanoparticles is approximately 17 nm.

Therefore, the diameter obtained by the experimental magnetic data was of 17 nm. This value is close to that (13 nm) calculated using the TEM images, the microscope used was a FEI Tecnai G2 Spirit Twin model, up to 120 kV with LaB6 (Lanthanum hexabrid) filament the images were obtained at 120 kV (see Fig. 8.(a) and Fig. 8.(b)), and considerably different from that (4.2 nm) obtained by XRD (Fig. 8.(d)) using a PANAlytical diffractometer, model X'PERT-PRO. Particle size distribution of the magnetic nanoparticles was also estimated by DLS the equipment is of the brand Horiba, model Nanopartica SZ-100 with laser 10 mW with incident beam with wavelength of 532 nm. The particle medium size is 38 nm (Fig. 8.(c)). As expected, the nanoparticle diameter obtained by DLS was larger than the other techniques, because the DLS technique measures the hydrodynamic diameter of the particles in the suspensions.

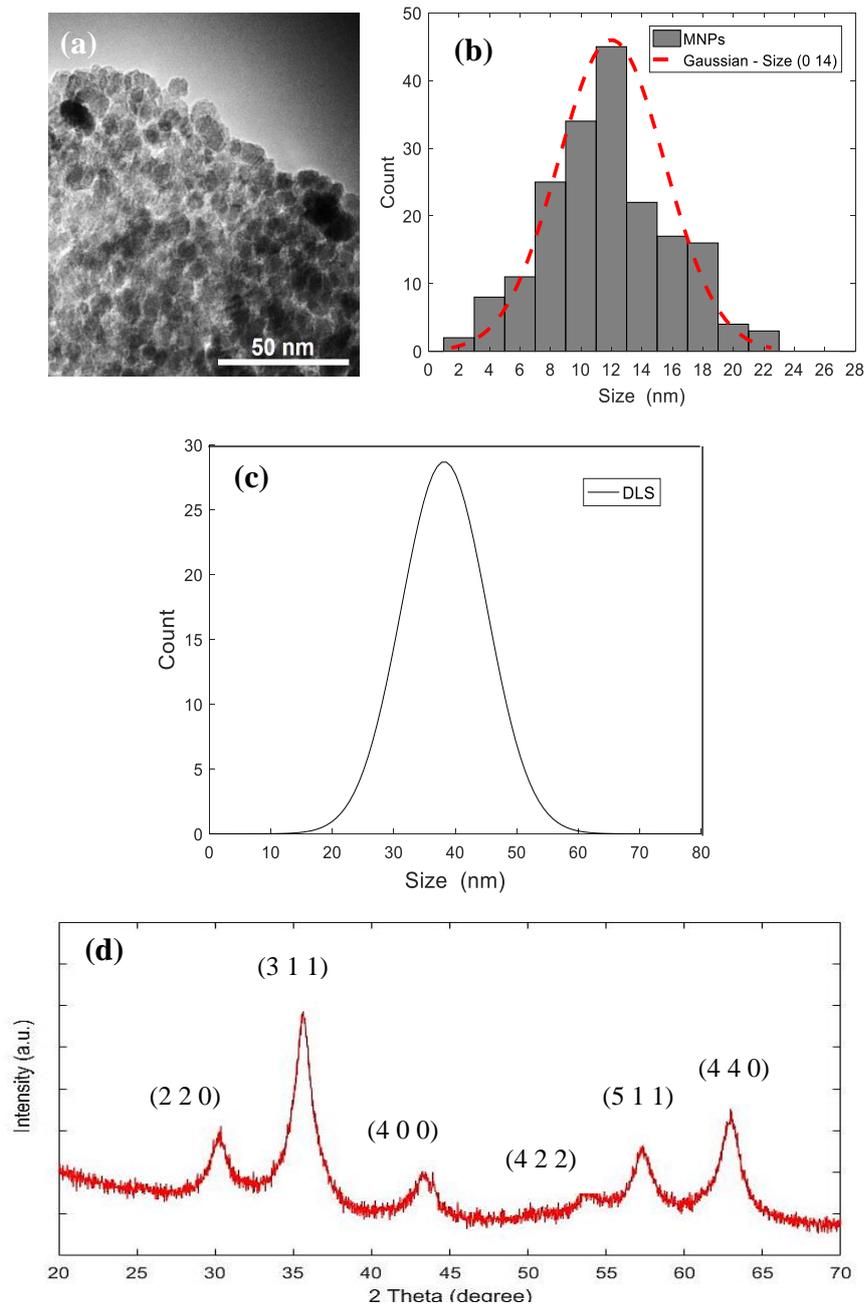

**Fig. 8.** (a) TEM image of the nanoparticles. (b) Histogram of the size distribution of the cobalt ferrite nanoparticles. (c) Particle size distribution of magnetic nanoparticles obtained by the DLS technique. (d) XRD of the cobalt ferrite nanoparticles.

Table 2 summarizes the size of magnetic nanoparticles measured by all techniques. The differences can be explained by the intrinsic characteristics of the analysis method [23].

Table 2. Comparison of the particle sizes obtained by various techniques.

| Technique | Size |
|-----------|------|
| TEM | 13 nm |
| DLS | 38 nm |
| XRD | 4.2 nm |
| Magnetic | 17 nm |

Hence, our magnetometer can used to obtain the magnetic characteristics of the samples and estimate the average size of nanoparticles with a reasonable accuracy, which is of practical interest. Additionally, this method is significantly less expensive than the TEM technique, enabling more laboratories to participate in advanced studies on these materials.

**5. Conclusions**

The magnetic measurement results of the nanoparticles and other samples obtained with the magnetometer developed in this study were consistent and statistically comparable with those obtained using other instruments, such as the commercial magnetometers (SQUID and VSM). In all the investigated cases, the errors inherent in the measurement results were smaller than 2%. Our results showed that the low cost, versatile, and automated self-built magnetometer with Hall-effect sensors is an excellent alternative to perform rapid and straightforward magnetic characterizations at different samples and has a magnetic moment sensitivity of $1.3 \times 10^{-9}$ Am$^2$ to sensitivity of 493 mV/mT, the sensitivity too can be adjustable in the range of 10 to 493 mV/mT. Furthermore, the magnetic nanoparticles were uniform in terms of composition and size, according to the XRD and TEM analyses. These analyzes also showed that the characterization through the magnetic curves provided a good value for the average particle size. Hence, our Hall magnetometer can be utilized to characterize such particles in practical applications.

**Acknowledgments**

This study was funded in part by the following Brazilian agencies: The National Council for Scientific and Technological Development (Conselho Nacional de Desenvolvimento Científico e Tecnológico – CNPq), the Coordinating Agency for the Improvement of Higher Education (Coordenação de Aperfeicoamento de Pessoal de Nível Superior – Capes) and the Research Support Foundation of Rio de Janeiro (Fundação de Amparo à Pesquisa do Estado do Rio de Janeiro – FAPERJ). We thank Professors R.L. Sommer of CBPF, N. Massalani, and M.A. Novak of UFRJ for the measurements with the VSM and SQUID magnetometers. We thank to Prof. Sonia Letichevsky for the XRD analysis of the samples, João Manuel for building the integrated circuit board and João Marcelo de Mendonça for the assistance in drawing Fig. 1 in AutoCAD®.